\documentclass[%
 aps, superscriptaddress,
 amsmath,amssymb,
 reprint, %
]{revtex4-1}

  \usepackage[dvips]{graphicx}

\usepackage{dcolumn}
\usepackage{amsmath}
\usepackage{latexsym}
\usepackage{color}
\usepackage{ulem}
\usepackage{verbatim}

\begin{document}

\title{Superconducting microwave resonators with non-centrosymmetric nonlinearity }

\author{M. Khabipov}
\affiliation{Physikalisch-Technische Bundesanstalt, Bundesallee
100, 38116 Braunschweig, Germany}
\author{V. Gaydamachenko}
\affiliation{Physikalisch-Technische Bundesanstalt, Bundesallee
100, 38116 Braunschweig, Germany}
\author{C. Kissling}
\affiliation{Physikalisch-Technische Bundesanstalt, Bundesallee
100, 38116 Braunschweig, Germany}
\affiliation{Advanced Electromagnetics Group, Technische Universit\"{a}t
Ilmenau, 98693 Ilmenau, Germany}
\author{R. Dolata}
\affiliation{Physikalisch-Technische Bundesanstalt, Bundesallee
100, 38116 Braunschweig, Germany}
\author{A. B. Zorin}
\affiliation{Physikalisch-Technische Bundesanstalt, Bundesallee
100, 38116 Braunschweig, Germany}

\date{April 12, 2022}

\begin{abstract}
We investigated both theoretically and experimentally open-ended coplanar waveguide
resonators with rf SQUIDs embedded in the central conductor at different positions.
These rf SQUIDs can be tuned by an external magnetic field and thus may exhibit the
non-centrosymmetric nonlinearity of $\chi^{(2)}$ type with suppressed Kerr nonlinearity.
We demonstrated that this nonlinearity allows for efficient mixing
of $\lambda/2$ and $\lambda$ modes in the cavity
and thus enables various parametric effects
with three wave mixing. These effects are the second harmonic generation, the half
tone generation, the parametric amplification in both degenerate and non-degenerate
regimes and deamplification in degenerate regime.


\end{abstract}
\maketitle

\section{Introduction}

Superconducting coplanar waveguide (CPW) resonators with embedded
Josephson junctions and SQUIDs are widely used for parametric
amplification \cite{Castellanos-Beltran2007,Yamamoto2008,Eichler2011},
bifurcation-based quantum detection
\cite{Manucharyan2007,Metcalfe2007,Vijay2009,Tancredi2013,Wustmann2013},
generation of nonclassical states of
microwaves \cite{Castellanos-Beltran2008,Wustmann2013,Zhong2013,Schneider2018},
studying the dynamical Casimir effect and photon field correlations
\cite{Wilson2011,Lahteenmaki2013}, parametric down conversion \cite{Chang2020}, etc.
At the working microwave frequencies (up to approximately 20~GHz), superconducting
resonators have very low losses, while the Josephson tunnel junctions
facilitate the parametric effects.
The operation of these circuits is usually based either on the Kerr
nonlinearity of the Josephson inductance
\cite{Castellanos-Beltran2007,Metcalfe2007,Vijay2009,Tancredi2013,
Castellanos-Beltran2008,Palacios-Laloy2008,Bourassa2012,Wustmann2013,Eichler2014}
or on a periodic modulation of the dc SQUID inductance by means of an
alternating magnetic flux \cite{Yamamoto2008,Schneider2018,Wilson2011,Lahteenmaki2013}.

Recently, the toolbox of superconducting quantum technologies \cite{Devoret2013}
has been supplemented with the elements having non-centrosymmetric
nonlinearity of $\chi^{(2)}$
type \cite{Shen1984}. These superconducting elements are based either on rf SQUIDs
\cite{Yurke1989,Zorin2016,Zorin2017}, asymmetric dc SQUIDs \cite{Chang2020}, or
the multijunction SQUID, i.e., the so-called superconducting nonlinear
asymmetric inductive elements (SNAILs) \cite{Zorin2017,Frattini2017}.
For the optimal constant flux $\Phi_e$ applied to the SQUID loop,
the current-phase relation in these elements may have the Kerr-free
shape \cite{Zorin2016,Zorin2017,Frattini2017},
\begin{equation}
I(\varphi) \approx (\varphi - \beta \varphi^2)\varphi_0 L^{-1}_s,
\label{beta-nonlinearity}
\end{equation}
where $\varphi_0 = \Phi_0/2\pi$ is the reduced magnetic flux quantum,
$L_s$ is the linear SQUID inductance, and nonlinear coefficient $\beta$ is the
electrical analog of susceptibility tensor $\chi^{(2)}$ in optics \cite{Shen1984}.
The nonlinear relation (\ref{beta-nonlinearity}) enables three wave mixing (3WM)
including frequency doubling and parametric down-conversion.
Optical crystals having nonzero susceptibility $\chi^{(2)}$
are quite rare in nature \cite{Kurtz1968} and fabrication of optical fibers with
non-centrosymmetric nonlinearity suitable for engineering traveling wave amplifiers
is not as easy as fabrication of the silica fibers with Kerr nonlinearity \cite{Agrawal}.
However, the superconducting technology allows the fabrication of the transmission
lines with embedded rf SQUIDs. These circuits can have a nonlinearity given
by Eq.(\ref{beta-nonlinearity}) and, thus, enable parametric amplification
of traveling microwaves using 3WM \cite{Zorin2016,Zorin2017,Miano2018}.
A recent study of the effect of parameter variations \cite{Peatain2021} confirmed
that such a parametric amplifier is a strong candidate for achieving a high
gain in a wide frequency band together with a high fabrication yield.

Parametric amplification based on serial arrays of SNAILs inserted
in the CPW resonators was recently demonstrated by the Yale group.
These Josephson parametric amplifiers with 3WM clearly demonstrated
a number of advantages of the Kerr-free operation including an improved dynamic
range  \cite{Sivak2019}, relatively large saturation power with a widely tunable
bandwidth  \cite{Frattini2018,Miano2021}, and near-quantum-limited performance \cite{Sivak2020}.

In this paper we first examined Nb open-ended CPW resonators with rf SQUIDs
embedded in their center, i.e. in the antinode of the fundamental ($\lambda/2$) mode.
The parameters of rf SQUIDs were close to those of the rf SQUIDs exploited
in the Josephson traveling wave parametric amplifiers (JTWPA) with 3WM \cite{Zorin2017}.
Our primary motivation was the investigation of these tunable nonlinear elements,
including the determination of their electric parameters.
These parameters were found from
the dependence of the resonant frequency on magnetic flux $\Phi_e$.

The tunability of the rf SQUID inductance has been utilized earlier in microwave
circuits for the coupling of, for example, two resonators \cite{Wulschner2016}
or a resonator and a phase qubit \cite{Allman2014}.
Here we investigate the circuit with the intermode coupling based on the
nonlinear characteristic of the rf SQUID given by Eq.~(\ref{beta-nonlinearity}).
For this purpose we designed CPW resonators with the rf SQUID
positioned at one third of the open-ended resonator length.
Thus, we engineered an artificial medium with a nonlinearity of $\chi^{(2)}$
type enabling efficient coupling of primarily the $\lambda/2$ and $\lambda$ modes.
In this circuit, we demonstrated a number of parametric 3WM phenomena including
the second harmonic generation (SHG), the half tone generation (HTG),
the parametric amplification in both degenerate and non-degenerate regimes, etc.
The improved design of a circuit with a high degree of intermode coupling
is proposed.

\section{Design and fabrication}

The design of our open-ended superconducting CPW resonators
(the microwave analog of a Fabry-Perot
cavity) with an embedded rf SQUID is schematically shown in Fig. 1.
The resonant frequencies of $\lambda/2$ mode were designed to be around 3.9 GHz.
The resonator length was 16.136 mm with a center-conductor width
of $w=32~\mu$m and gaps between this conductor and the ground
plane conductors of $s=16~\mu$m. The nominal thickness of Nb layer was 200 nm.
These dimensions of the CPW waveguide
yielded a wave impedance of $Z_0 = 50~\Omega$.
This impedance was matched with the impedance of the input and output lines.
The CPW waveguide parameters were chosen similar to those of the CPW transmission
line used in the design of the JTWPAs \cite{Zorin2017}.
The total inductance and the total capacitance of the resonator were
$L_{\textrm{tot}} = \ell d = 6.454$~nH and $C_{\textrm{tot}} = c d = 2.582$~pF,
respectively. The corresponding specific values were $\ell =0.4$~pH/$\mu$m
and $c = 0.16$~fF/$\mu$m.

The input, $C_{\textrm{in}}$, and output, $C_{\textrm{out}}$, capacitances of
the resonator were realized either as gap capacitances with spacing $w_g= 2~\mu$m
or interdigital capacitances having 4 fingers (see, e.g., Ref. \cite{Goeppl2008}).
The finger width and spacing between the fingers were 2~$\mu$m, while
the finger length varied from 10~$\mu$m to 50~$\mu$m.
The nominal capacitance values $C_\textrm{in}$ and $C_\textrm{out}$ for
Sample 1 were 1.8~fF and 2.8~fF, respectively, and 2.8~fF and 14~fF for
Sample 2, respectively.
In the case of Sample 1 these values result in a coupling quality factor
\cite{Goeppl2008} $Q_c=94\,000$ for the fundamental $\lambda/2$ mode, while
the loaded quality factor $Q \ll Q_c$.
The experimental value $Q\approx 4\,000\,...\,8\,000$
was increasing for an increasing drive power
$-105\,$dBm$\,...\,-90\,$dBm, what can be explained by the saturation of microscopic
two-level systems \cite{Martinis2005,Burnett2016} and indicates that $Q$ is mainly
determined by the dielectric loss in the deposited SiO$_2$ layer \cite{deGraaf2020}
and/or in the Josephson junction barriers \cite{Weides2011}.

Sample 2 was designed to be critically coupled with $Q_c=5\,100$,
approximately equal to the internal quality factor, which is an optimum in
the compromise of coupling strength and quality factor needed for pronounced
nonlinear interactions \cite{Goeppl2008}. For this sake, we created a strongly
coupled output port, i.e., $C_\textrm{out}$ was chosen much larger
than $C_\textrm{in}$ \cite{Wustmann2013}. The resulting experimental quality
factor was $Q\approx 2\,000$ and didn't
exhibit significant temperature dependence in the range from 20~mK up to 4.2~K.

A local magnetic field was applied to the  rf SQUID loops via a control-current
line (the thin Nb wire seen very close to the upper ground plane in Fig.1c).
The thin-film rf SQUID inductances had the meander shape with a few turns
(see Fig.1c) giving the nominal values of $L_g$ around 30~pH.

The samples were fabricated using the Nb-trilayer technology with a critical
current density $j_c$ from 200 A$/$cm$^2$ to 500 A$/$cm$^2$ on a Si/SiO$_2$
substrate \cite{Dolata2005}.
The self-capacitance $C_J$ of the Josephson junctions with a nominal area of
1~$\mu$m$^2$ was in the range from $45-50$~fF.
The details of manufacturing similar CPW resonators with embedded Josephson
junctions and dc SQUIDs were described in Ref.~\cite{Khabipov2014}.

\section{Linear regime. Characterization of the rf SQUIDs}

\begin{figure}[t]
\begin{center}
\includegraphics[width=3.4in]{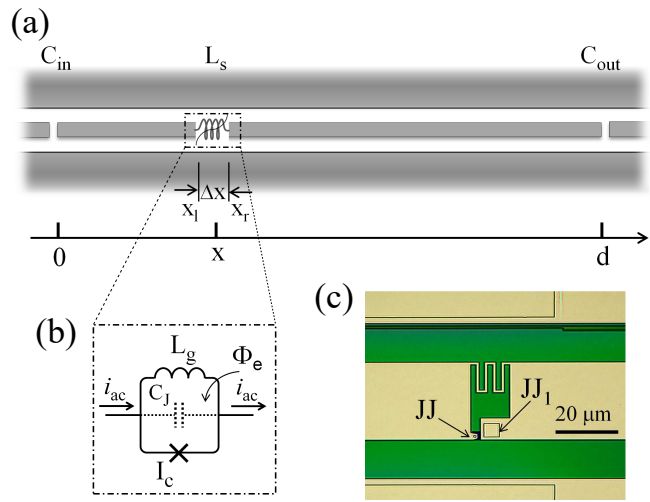}
\caption{(a) Schematic of the open-ended coplanar transmission line of total
length $d$ including inductance $L_s(\Phi)$ of physical length $\Delta x$ embedded
at position $x$. Electrical diagram (b) and microphotograph (c) of the Nb rf SQUID,
where an auxiliary (large) Josephson junction (JJ$_1$) in series with the primary (small)
junction (JJ) was added as a stud-via.}
\label{CPW-Line}
\end{center}
\end{figure}

The inverse inductance for a small alternating current
$i_\textrm{ac}$ (see Fig. 1b) passing through
the rf SQUID is given by the expression
\begin{equation}
L_s^{-1}[\Phi_{\textrm{dc}}(\Phi_e)]
= L_g^{-1} \left\{1 + \beta_L\cos[\Phi_{\textrm{dc}}(\Phi_e)/\varphi_0]\right\}.
\label{L_J}
\end{equation}
Here we assumed that the Josephson tunnel junction in the SQUID loop has
the sinusoidal (conventional) current-phase relation.
The constant flux in the SQUID loop $\Phi_{\textrm{dc}}$ is found by solving the
transcendental equation \cite{KK-book1986}
\begin{equation}
\Phi_e = \Phi_{\textrm{dc}} + \beta_L \varphi_0\sin(\Phi_{\textrm{dc}}/\varphi_0),
\label{phi-phie}
\end{equation}
where $\Phi_e$ is applied magnetic flux, $\beta_L = L_g I_c /\varphi_0 < 1$
is the dimensionless screening parameter,
$L_g$ is the SQUID inductance, and $I_c$ is the Josephson critical current.
Embedding the rf SQUID in the CPW resonator causes a shift of the resonant
frequency. The flux dependence of the rf SQUID inductance given
by Eq.~(\ref{L_J}) allows the characterization of the rf SQUID by measuring
that frequency shift. This method is conceptually similar to that developed
by Rifkin and Deaver, Jr. \cite{Rifkin1976} for the characterization of rf SQUIDs
having inductive coupling to a rf-driven tank circuit \cite{Ilichev2001}
(see also Ref.~\cite{Wegner2021}).

The resonant frequency of the $n$-th mode, $n\lambda/2 =d$, $n = 1, 2,...$,
is given by the general formula (see Eqs. (\ref{delta-fnRes})
and (\ref{ref-freq}) of Appendix):
\begin{equation}
 \frac{\omega_n}{\omega_n^{(0)}}
 =  1- \frac{L_s(\Phi_{\textrm{dc}})}{L_\textrm{tot}}
 \sin^2 \frac{n\pi x}{d}+\frac{\Delta x}{d}.
\label{f_res0}
\end{equation}
For the first mode, $n=1$, and the central
position of the rf SQUID, $x = d/2$, expression (\ref{f_res0}) takes the form
\begin{equation}
 \frac{\omega_1}{\omega_1^{(0)}} \approx  1- \frac{L_g/L_{\textrm{tot}}}
{1+\beta_L \cos(\Phi_{\textrm{dc}}/\varphi_0)},
\label{f_res}
\end{equation}
where we used Eq.~(\ref{L_J}) and neglected the small constant
term $\Delta x/d$ $\approx\,10^{-3}$
because of the small SQUID size $\Delta x$.

The bare resonant frequency in Eq.~(\ref{f_res}) is \cite{Pozar1993}
\begin{equation}
\omega_n^{(0)} = \frac{1}{\sqrt{L_{\textrm{eff},n} C_{\textrm{eff},n}}}
= \frac{ n \pi}{\sqrt{L_{\textrm{tot}}C_{\textrm{tot}}}},
\label{fRes0}
\end{equation}
where the effective $LC$ parameters for the $n$-th
mode are given by the following expressions \cite{Wallquist2006}:
\begin{equation}
\frac{1}{L_{\textrm{eff},n}} = \frac{(k_nd)^2 }{2 L_{\textrm{tot}}}
 \left[1+\frac{\sin 2k_n d}{2k_n d}\right] =\frac{ (n\pi)^2}{2 L_{\textrm{tot}}}
\label{Leff}
\end{equation}
and
\begin{equation}
C_{\textrm{eff},n} = \frac{C_{\textrm{tot}}}{2}
\left[1+\frac{\sin 2k_n d}{2k_n d}\right] = \frac{C_{\textrm{tot}}}{2}
\label{Ceff}
\end{equation}
with $k_n=k^{(0)}_n= n\pi/d$.
As long as the rf SQUID plasma frequency $\omega_J$ is sufficiently
high, that is $\omega_J = 1/\sqrt{L_g C_J}\approx 2\pi \times 120$~GHz
$ \gg \omega_n^{(0)} $, the effect of the Josephson junction self-capacitance
$C_J$ on the resulting resonant frequency $\omega_n$ can safely be
neglected at least for the first 10 modes.

\begin{figure}[t]
\begin{center}
\includegraphics[width=3.3in]{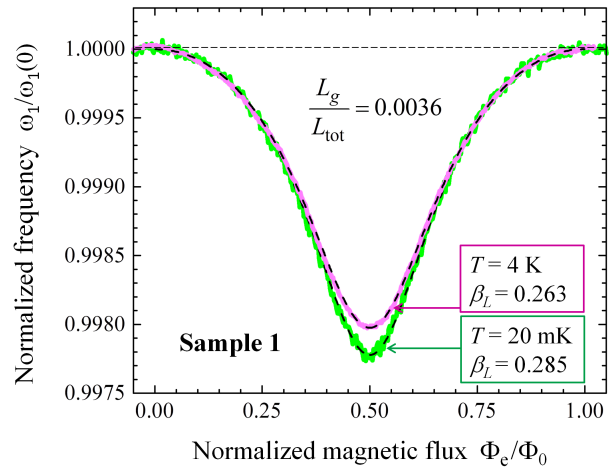}
\caption{One period of the normalized dependence of the resonant
frequency of the fundamental $\lambda/2$-mode, $\omega_1(\Phi_e)/\omega_1(0)$,
on the normalized magnetic flux, $\Phi_e/\Phi_0$, measured at two temperatures
in Sample 1.
The resonant frequency is normalized to its maximum value of
$\omega_1(0) = \omega_1^{(0)}\{1-L_g/[(1+\beta_L)L_\textrm{tot}]\}$.
In this sample, the rf SQUID is embedded in the center of the CPW
resonator, $x = 0.5d$.
Dashed lines denote the fitting curves using formula (\ref{f_res}).
Both measurements were performed in a dilution refrigerator.}
\label{Res-curves}
\end{center}
\end{figure}

The measurements of the resonant frequency of the fundamental
mode ($n = 1$) were performed (see Appendix C for details) in a dilution
refrigerator at temperatures $T = 20$~mK $\pm$ 0.2~mK
and $T = 4$~K $\pm$ 0.1~K.
The resonant frequency showed a clear periodic dependence on
the control current producing the magnetic flux $\Phi_e$.
The experimental data and the fits using formula (\ref{f_res}) are
presented in Fig.~2.
Because the values of the resonant frequency $\omega_1(0)$ found
in both experiments at zero magnetic flux $\Phi_e$ were identical (3.917~GHz),
we concluded that resonator inductance $L_\textrm{tot}$ doesn't depend on
temperature.

\begin{figure}[b]
\begin{center}
\includegraphics[width=3.3in]{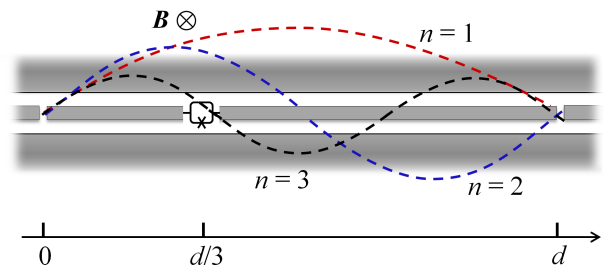}
\caption{Geometry of the open-ended CPW resonator with rf SQUIDs
embedded at $x = d/3$, yielding maximum coupling
of the first two resonant modes.
Dashed lines show schematically the profiles
of standing waves of the current for modes $n=1$, $n=2$, and $n=3$.}
\label{diagram-embedded}
\end{center}
\end{figure}

The participation ratio value, $L_g/L_\textrm{tot} = 0.0036$, found from
fitting the experimental curves turned out to be also temperature independent.
The values of the SQUID-parameter $\beta_L$ for different temperatures,
$\beta_L^{(4 \textrm{K})} = 0.263$ and $\beta_L^{(20 \textrm{mK})} = 0.285$
(see Fig.~2), clearly pointed to a temperature dependence of the critical
current. Thus, the ratio of the critical current values at the two
temperatures is
\begin{equation}
\iota =
\frac{I_c^{(20 \textrm{mK})}}{ I_c^{(4 \textrm{K})}}
= \frac{\beta_L^{(20 \textrm{mK})}}{\beta_L^{(4 \textrm{K})}}
\approx 1.08.
\label{beta-ratio}
\end{equation}
Using the Ambegaokar-Baratoff formula \cite{AB1963} for the temperature
dependence of the critical current of an ideal
tunnel junction between two identical BCS superconductors
with critical temperature $T_c =T_c^\textrm{(Nb)} \approx$ 9.0~K (here
$T_c^\textrm{(Nb)}$ is the critical
temperature of our Nb films) we arrive at the same value
of $\iota_\textrm{AB} = 1.08$.

\section{Nonlinear effects}

Setting the external magnetic flux $\Phi_e$  in Eq.~(\ref{phi-phie}) such
that the constant flux $\Phi_{\textrm{dc}} = \Phi_0/4$, that is
\begin{equation}
\Phi_e/\varphi_0 = \pi/2 + \beta_L,
\label{Phi-Kerr-free}
\end{equation}
yields the Kerr-free nonlinear inductance of the rf SQUID of the form given
by Eq.~(\ref{beta-nonlinearity}) \cite{Zorin2016}. The nonlinear coefficient in
the current-phase relation (\ref{beta-nonlinearity}), given in the general case by
\begin{equation}
\beta = 0.5\beta_L \frac{\sin (\Phi_{\textrm{dc}}/\varphi_0)}
{1+\beta_L \cos (\Phi_{\textrm{dc}}/\varphi_0)},
\label{beta-Kerr-free-value}
\end{equation}
then equals $\beta=0.5 \beta_L$. For $\Phi_{\textrm{dc}} = \Phi_0/4$,
the linear inductance of the rf SQUID (\ref{L_J}) is equal to
its geometrical inductance, $L_s = L_g$, while the inductance of the Josephson
junction is infinite,
$L_J = \varphi_0/[ I_c\cos(\Phi_{\textrm{dc}}/\varphi_0)] \rightarrow \infty$.

\begin{figure}[b]
\begin{center}
\includegraphics[width=3.45in]{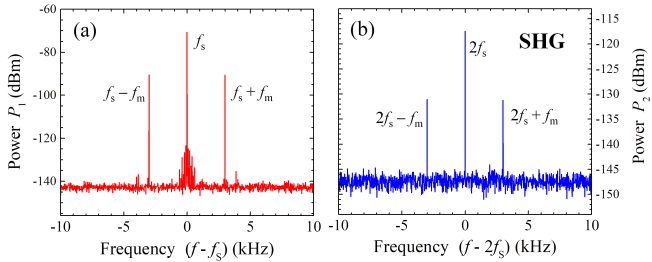}
\caption{SHG in Sample 2 ($x=d/3$ and $\beta_L = 0.44$).
(a) Spectrum of the amplitude-modulated signal
(\ref{AM-input-signal-cavity}) with carrier frequency of
$f_s=\omega_s/2\pi$ = 3.8858~GHz $\approx \omega_1/2\pi$
= 3.888 GHz,
modulation frequency of $f_m=\omega_m/2\pi$ = 3~kHz, and
modulation index of $\mu = 0.2$.
(b) Spectrum of the generated output signal having the form of
a triplet with the central line at the double signal
frequency, $2f_s = 7.7716$~GHz, and two sidebands at $2f_s\pm f_m$.
The drive signal power is $P_s = -87$~dBm.
The measurements were performed in a helium bath at temperature
$T=4.2$~K.}
\label{SHG}
\end{center}
\end{figure}

\begin{figure}[b]
\begin{center}
\includegraphics[width=3.45in]{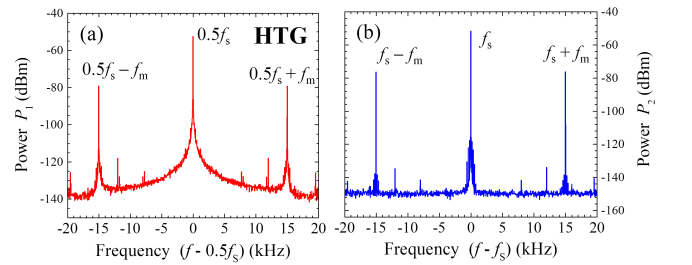}
\caption{(a) Spectrum of the output signal appeared due to HTG
in Sample 2 ($x=d/3$ and $\beta_L = 0.44$).
(b) Spectrum of the amplitude-modulated signal with carrier
frequency of $f_s=$ 7.76964 GHz $\approx \omega_2/2\pi$ = 7.737 GHz
and modulation frequency of $f_m=15$~kHz applied to the resonator.
The spectrum of the generated half tone (a) includes
a central peak at frequency $f = f_s/2 = 3.88482$~GHz
and two sidebands at frequencies $ f_s/2 \pm f_m$.
The signal power is $P_s = -52$~dBm.
The measurements were performed in a helium bath at
temperature $T=4.2$~K.}
\label{HTG}
\end{center}
\end{figure}

To realize efficient mixing of the two lowest modes, we moved the
rf SQUID out of the resonator center, because at $x=0.5d$ mode $n=2$ has
a node in the standing-wave of current and, hence, is only weakly coupled
to fundamental mode $n=1$. Embedding the rf SQUID at $x = d/3$ yields nonzero
values of $|\sin(n\pi x/d)| =|\sin(n\pi /3)|= \sqrt{3}/2$, for all integer
$n$ except $n= 3, 6, 9,...$. Then the three-photon coupling coefficient
(see Eq.~(\ref{coupling-coeff}) of Appendix B) is
\begin{eqnarray}
B_{lmn} &\propto& \beta \sin k_{l}x \, \sin k_m x \, \sin k_{n}x \nonumber\\
&\approx& \beta \sin (l\pi /3) \sin (m\pi/3) \sin (n\pi/3).
\label{coupling3modes}
\end{eqnarray}
For modes $l=1$, $m=1$, and $n=2$, the value of $|B_{112}|$
is maximum, while, for example, for modes $l=1$, $m=2$, and $n=3$,
its value is zero, $|B_{123}| = 0$.
The latter property is evident, because the rf SQUID position, $x=d/3$,
coincides with the standing-wave node of mode $n =3$ (see the corresponding
dashed curve in Fig.~3).

\subsection{Second harmonic generation}

Up-conversion (doubling) of the frequency of a light beam while preserving its
quantum state \cite{Huang1992} is of particular importance in nonlinear optics.
A laser beam in this process is usually passing through a large crystal
having non-centrosymmetric nonlinearity \cite{Boyd2008}. Using of an optical cavity
containing a nonlinear crystal and resonant at the second harmonic may enable
a source of ultraviolet radiation within this cavity \cite{Wu1985}.
To demonstrate the intracavity up-conversion of microwaves we applied to the circuit
a high frequency drive signal
\begin{equation}
I_s(t) = I_{s0}(t)\sin\omega_s t
\label{AM-input-signal}
\end{equation}
with a slowly oscillating amplitude having the shape
\begin{equation}
I_{s0}(t) = A_0(1+\mu \sin\omega_m t),
\label{AM-input-signal-Apm}
\end{equation}
where carrier frequency $\omega_s \approx \omega_1$, modulation
frequency $\omega_m \ll \omega_{1,2}/Q_{1,2} \ll \omega_1$, $A_0 = \textrm{const}$,
and modulation index $\mu \ll 1$.
For sufficiently small driving signal (\ref{AM-input-signal}), the fundamental-mode
oscillations in the cavity have a shape similar to that of the steady-state
oscillations in a driven oscillator (see, for example, Ref. \cite{Migulin1983}),
\begin{equation}
I_1 = I_{0}(t) \sin(\omega_s t + \delta + \delta_0)
\label{AM-input-signal-cavity}
\end{equation}
with the amplitude
\begin{equation}
I_0(t) \propto \frac{I_{s0}(t)}{\sqrt{[1- (\omega_s/\omega_1)^2]^2
+ (\omega_s/\omega_1)^2/Q_1^2}}
\label{amp-cavity}
\end{equation}
and phase $\delta$, that is
\begin{equation}
\tan\delta = \frac{\omega_s\omega_1}{Q_1(\omega_s^2-\omega_1^2)}.
\label{phase-cavity}
\end{equation}
Constant phase $\delta_0$ is determined by
the parameters of the measuring setup.

The power spectra measured in the vicinities of frequencies $\omega_s$ and
$2\omega_s$ are shown in Fig.~4. One can see the generated frequency triplet
consisting of the double frequency carrier at $\omega = 2\omega_s$ with power
$P_2$ and two sidebands at frequencies $2\omega_s\pm \omega_m$ with powers
$(\mu^2/4)P_2$ (see panel (b)). In the time domain, this signal has the shape
\begin{equation}
I_2 \propto (1+\mu \sin\omega_m t) \sin (2\omega_s t + \vartheta),
\label{output-SHG}
\end{equation}
where $\vartheta$ is the relative phase of the generated signal which is
coupled to the total phase of the fundamental mode, $\delta + \delta_0$.
The relatively small output power $P_2$ can be explained by rather large
frequency mismatch, $(2\omega_s-\omega_2)/2\pi = 37.1$~MHz
$\gg \omega_2/(2\pi Q_2) \approx 3.86$~MHz,
thus the generated signal is off-resonant for the second mode.

\begin{figure}[b]
\begin{center}
\includegraphics[width=3.0in]{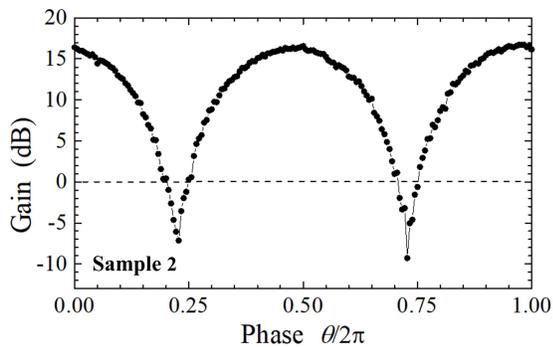}
\caption{Phase-sensitive parametric amplification and deamplification
in Sample 2 (Fig.~3). The input
signal frequency and the power are $\omega_s/2\pi$ = 3.875~GHz and $-140$~dBm,
respectively. The pump frequency
and the power are $\omega_p/2\pi$ = 7.75~GHz and $-70$~dBm, respectively.
The relative phase between these two phase-locked signals is $\theta$ (\ref{degenerJPA}).
Dashed line shows the level of the output signal power when the pump is off.
The maximum signal gain is 17 dB, while the maximum deamplification is about $-8$ dB.
The measurements were performed in a dilution refrigerator at temperature $T=18$~mK.}
\label{deAmp}
\end{center}
\end{figure}

\begin{figure}[b]
\begin{center}
\includegraphics[width=3.2in]{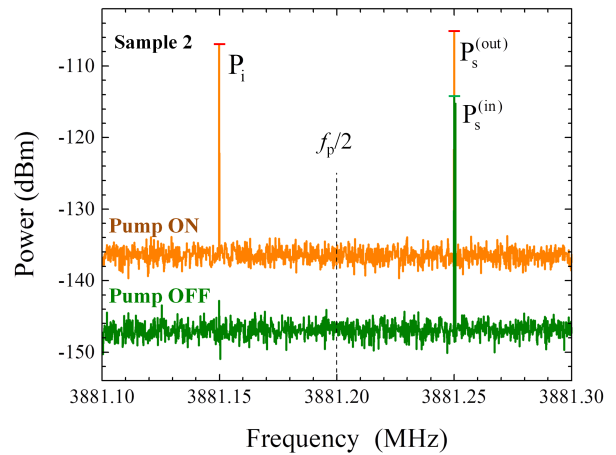}
\caption{Parametric amplification of signal tone $f_s = 3.88125$~GHz
by means of 3WM in the non-degenerate case (pump frequency $f_p=7.7624$~GHz and
idler frequency $f_i=f_p - f_s=3.88115$~GHz).
The pump power is $P_p \approx-65$ dBm.
The measurements were performed in a helium bath at $T=4.2$~K.}
\label{JPA-non-degenerate}
\end{center}
\end{figure}

\subsection{Half tone generation}

Applying an intensive harmonic signal with frequency $\omega_s$ close to the double
resonant frequency of the fundamental mode, $2 \omega_1 \approx \omega_2$,
may result in HTG, or, equivalently, the oscillation
period doubling. (Note that the period tripling was earlier observed
in a superconducting resonator with the Kerr nonlinearity
by Svensson et al. \cite{Svensson2017}.)
The period doubling
is possible within a finite range of signal frequencies
$\omega_s$ around $2\omega_1$, i.e.,
$ \delta\omega \leq |\omega_s/2 - \omega_1|$.
Frequency range $\delta\omega$ depends on
the losses for the half tone mode and the signal power that should
be sufficient for compensating these losses \cite{Migulin1983}.
In this case, the zero state of mode $\omega_1$ becomes unstable and
the circuit switches in one of the oscillating states, both with frequency
$\omega_s/2$ and the relative phase difference of $\pi$ \cite{Zorin2011}.

Figure \ref{HTG}a shows HTG in the case of an amplitude-modulated input
signal (\ref{AM-input-signal}) with  $\omega_s \approx 2\omega_1$,
or $\delta\omega \approx 0$,
whose power spectrum is presented in Fig.~\ref{HTG}b.
The output spectrum measured around the half signal frequency,
$\omega \sim 0.5\omega_s$  (shown in panel (a))
mimics the spectrum of the input signal (panel (b)).
It consists of the carrier frequency $(= 0.5\omega_s)$ and two sideband
peaks. Due to a very small modulation frequency
$(\omega_m \ll \omega_{1,2}/Q_{1,2})$ the input signal
can be considered as a harmonic signal with a slowly-varying amplitude.
Due to the down-conversion (coupling coefficient $B_{211} \neq 0$) and
a sufficiently small modulation index, $\mu \ll 1$, the output
signal at frequency $\omega =0.5\omega_s$ also has an amplitude which
is slowly-varying with frequency $\omega_m$.
Thus, its spectrum presents a triplet, where the small sideband peaks
are positioned at $0.5\omega_s \pm \omega_m$.

Relative phase $\vartheta$ of the generated
tone, $I_{\omega_s/2} \propto \sin(0.5 \omega_s t+ \vartheta)$, takes randomly
one of the two values, $\vartheta = \vartheta_{1,2}$. In the general case, these
values depend on the power ($\propto I_s^2$) and the dissipation in the circuit,
but always have a fixed difference, $| \vartheta_1 -\vartheta_2| = \pi$
(see, for example, Eq.~(19) in Ref. \cite{Zorin2011}).

\subsection{Parametric amplification}

To demonstrate the operation of the circuit in the regime
of a Josephson parametric amplifier (JPA) with 3WM in the degenerate mode,
we applied a small harmonic signal $I_s$ of frequency $\omega_s \approx \omega_1$
and large pump, $|I_p| \gg |I_s|$, at the double frequency,
$\omega_p =2\omega_s$, with relative phase difference $\theta$,
\begin{equation}
I_s + I_p =I_{s0} \sin(\omega_s t + \theta) + I_{p0} \sin 2\omega_s t.
\label{degenerJPA}
\end{equation}
Figure (\ref{deAmp})
shows the measured amplification/deamplification versus phase $\theta$.
The maximum gain of 17 dB is comparable with the figure reported
by Yamamoto et al. \cite{Yamamoto2008} for the flux-driven Josephson parametric
amplifier based on a $\lambda/4$-cavity terminated by a dc SQUID.
The 4WM amplifier of Castellanos-Beltran and Lehnert \cite{Castellanos-Beltran2007},
based on a serial array of dc SQUIDs embedded in a $\lambda/4$ CPW
resonator, showed a gain slightly above 20 dB.
The largest deamplification (about $-8$~dB) measured in our sample is weaker than
that achieved in Refs.~\cite{Yamamoto2008} and \cite{Zhong2013} (ca. $-20$~dB).
This may be associated with relatively large background noise in our experiment.
Still, the observed deamplification can be interpreted
as a fingerprint of background-noise squeezing
\cite{Yurke1989,Castellanos-Beltran2008}.

Using the obvious advantage of 3WM, that the frequencies of an intensive
pump and a small signal belong to different modes,
we have realized the non-degenerate mode of operation of our JPA.
The gain in this case is phase preserving. Fixing a small harmonic
signal of frequency $\omega_s\approx \omega_1$
slightly above the half of the pump frequency,
$(\omega_s - \omega_p/2)/2\pi = 50~\textrm{kHz} < \omega_1/Q$, signal gain
$G = P^{\textrm{(out)}}_s/P^{\textrm{(in)}}_s$ of about 9~dB was observed.
This direct gain was accompanied by the cross-gain
$G_\textrm{cross} =P_i/P^{\textrm{(in)}}_s$ (approximately 8~dB),
where the idler at frequency $\omega_i = \omega_p - \omega_s \lesssim \omega_s$
has, according to the Manley-Rowe relation \cite{Manley-Rowe1956},
the power $P_i \lesssim P^{\textrm{(out)}}_s$ (see Fig.~\ref{JPA-non-degenerate}).
Both the direct gain $G$ and the cross-gain $G_\textrm{cross}$ are proportional
to nonlinear coefficient $\beta^2$ and, hence, are sensitive to the setting
of magnetic flux $\Phi_e$.

The latter property together with the periodic dependence of the nonlinear
coefficient $\beta$ on the external flux $\Phi_e$ with the period of
$\Delta \Phi_e = \Phi_0$,
can be used for the evaluation of SQUID-parameter $\beta_L$.
Using relations (\ref{phi-phie}) and (\ref{beta-Kerr-free-value}) one can
find the values of magnetic flux $\Phi_e$ giving the maximum of $\beta^2$.
Within one period, $0\leq \Phi_e<\Phi_0$, these two values are
\begin{equation}
\Phi_{e1} =\varphi_0 \arccos(- \beta_L) + \varphi_0 \beta_L\sqrt{1-\beta^2_L}
\label{Phi-e1-max-b}
\end{equation}
and
\begin{equation}
\Phi_{e2} = \Phi_0 - \Phi_{e1},
\label{Phi-e2-max-b}
\end{equation}
respectively. Note, that the magnetic flux value given by
Eq.~(\ref{Phi-e1-max-b}) is somewhat larger than the value given
by Eq.~(\ref{Phi-Kerr-free}) for the Kerr-free case, although the
difference between these two values is vanishingly small for $\beta_L \ll 1$.
As long as flux $\Phi_e$ is proportional to the control current strength, the ratio
$(\Phi_{e2}-\Phi_{e1})/\Phi_0 = (\Phi_{e2}-\Phi_{e1})/(\Phi_{e2}+\Phi_{e1})$
can be found from the corresponding ratio of the control currents.
The above ratio allows finding parameter $\beta_L$.
Thus, $\beta_L$ found in Sample 2 from the parametric cross-gain measurements
is 0.446, while the value found from the resonance measurements of
this sample is $\beta_L = 0.44$. 

Finally, we note that our open-ended cavity can also operate as
a two-mode parametric frequency converter \cite{Migulin1983}.
(Compare with the ring modulator based on the Wheatstone bridge
configuration of four Josephson junctions enabling
mixing of the waves in three attached resonators \cite{Bergeal2010}.)
For example, applying a large pump $(\omega_p)$ and a small
signal $(\omega_s)$, both with frequencies around $\omega_1$,
gave rise to generating the sum frequency signal (SFG),
$\omega_\textrm{sum} =\omega_s+ \omega_p \approx \omega_2$.
Similarly, the difference frequency generation (DFG),
$\omega_\textrm{dif} =\omega_s- \omega_p \approx \omega_1$, was also
possible, where input signal frequency $\omega_s \approx \omega_2$,
while pump frequency $\omega_p \approx \omega_1$.

\section{Discussion and outlook}

\begin{figure}[t]
\begin{center}
\includegraphics[width=2.5in]{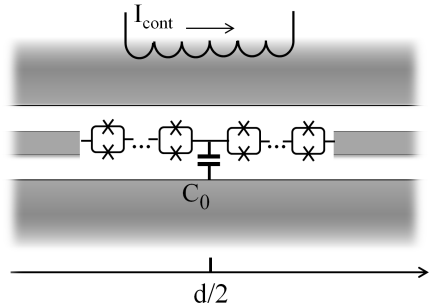}
\caption{Array of $N$ dc SQUIDs and ground capacitor $C_0$ embedded
in the central part of the CPW resonator, $x=d/2$.
Tuning the reduction of the resonant frequencies of the odd modes is possible
by means of constant control current $I_\textrm{cont}$ creating a local
magnetic field and thus varying the dc SQUID array inductance.
A necessary reduction of the resonant frequencies of the even modes occurs
due to the ground capacitance $C_0$.
}
\label{dcSQUIDinsert}
\end{center}
\end{figure}

In the open-ended cavity
with finite coupling capacitances $C_\textrm{in}$ and $C_\textrm{out}$
and nonzero participation ratio $\alpha$ the resonant frequencies
of the modes are not strictly equidistant \cite{Goeppl2008}.
One can, however, modify the circuit design such that
for modes $l=m=1$ and $n=2$ the ratio of the
resonant frequencies is 1:2. Then, 3WM condition,
$\omega_l+\omega_m-\omega_n=0$, is strictly fulfilled for the
first two resonant frequencies of the cavity.

A possible modification of the circuit is shown schematically
in Fig.~\ref{dcSQUIDinsert}, where
the array of $N$ dc SQUIDs \cite{Palacios-Laloy2008} and a relatively large ground
capacitance $C_0$ \cite{Winkel2020} are embedded in the center of the CPW resonator.
As long as the critical
current of the Josephson junctions in these dc SQUIDs is relatively large,
their Kerr nonlinearity is small. However, the SQUID chain
inductance and, thus, the resonant frequencies of all odd modes
including $n=1$ (having a current antinode at $x=d/2$) can be varied
in a wide range  with the help of dc control
current $I_\textrm{cont}$ \cite{Palacios-Laloy2008}.  In a similar
manner, ground capacitance $C_0$
reduces the resonant frequencies of the even modes (having a voltage antinode
at $x=d/2$). For a sufficiently large inductance of the dc SQUID chain and
a sufficiently large capacitance $C_0$, equidistance of all cavity modes with
$n \geq 3$ is destroyed, while the frequency ratio for the two lowest modes,
$\omega_2/\omega_1=2$, can be kept fixed by adjusting $I_\textrm{cont}$.

\begin{figure}[t]
\begin{center}
\includegraphics[width=3.0in]{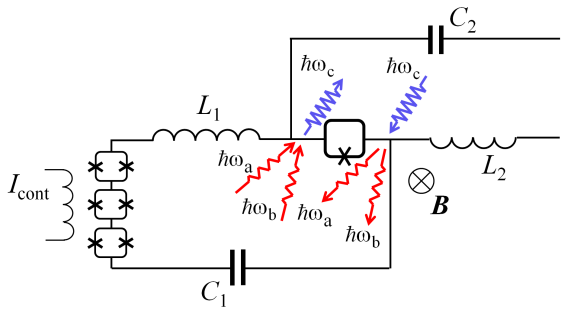}
\caption{Schematic of a lumped-element circuit comprising two resonators with
the ratio of resonant frequencies $\omega_2/\omega_1 =2$ and having nonlinear
coupling of $\chi^{(2)}$-type.
In the quantum case, this circuit enables SFG with single photons,
$\hbar\omega_a+\hbar\omega_b= \hbar\omega_c$, where input frequencies
$\omega_a\approx\omega_b~ (\approx\omega_1)$. Under a drive at frequency
$\omega_c\approx\omega_2$, a destruction of photon $\hbar\omega_c$ gives rise
to creation of the twin photons with frequencies
$\omega_a\approx\omega_b ~ (\approx\omega_1)$.
}
\label{Eqv-coupled-LCcircuits}
\end{center}
\end{figure}

The system of two modes tuned in the resonance of type $2\omega_1 = \omega_2$,
and having nonlinear coupling of $\chi^{(2)}$ type can be of particular
interest in physics \cite{Neyfeh1995}. Our CPW cavity with the improved design
or its lumped-element analog (see Fig.~\ref{Eqv-coupled-LCcircuits}) could be a
such a system implemented with the help of superconducting elements
and microwave frequencies.
For example, if energy losses are negligibly small and a drive is applied
to mode $\omega_1$ the energy may
be periodically transferred from mode $\omega_1$ to mode $\omega_2$ and vice versa.
This behavior mimics the beating-like dynamics of two weakly-coupled identical
pendulums \cite{Migulin1983}. However, there is a principal difference between
these two phenomena; in the case of the circuit with parametric
coupling the energy exchange may occur between two oscillators with
frequencies being in the ratio of two to one.
Similar behavior associated with nonlinear coupling may occur, for example,
with the pitch and roll modes of ship motions when the ratio of these
frequencies is close to two due to unequal moments of inertia for
these two modes \cite{Neyfeh1973}.

The superconducting cavity with the predominant coupling of modes
$n=1$ and $n=2$ allows for efficient generation of nonclassical states
of microwaves. Due to strict phase relations between the input and the output
signals, a coherent transfer of the source phase from
one part of the spectrum to another  is possible.
For example, SHG, SFG, and DFG, on the one hand, and HTG and JPA,
on the other hand, give rise to quantum frequency conversion
(QFC) \cite{Shen1984} and spontaneous parametric down-conversion
(SPDC) \cite{Klyshko1988}, respectively.

In the case of sufficiently low photon-loss rate, $\kappa \ll g$,
SFC (with quantum relation $\hbar\omega_a + \hbar\omega_b
= \hbar\omega_c$) and SPDC ($\hbar\omega_c = \hbar\omega_a + \hbar\omega_b$)
are described by the second quantization Hamiltonian (\ref{Ham112})
with the relevant part
\begin{equation}
H_\textrm{nl} = \hbar g(a_a a_b a_c^\dag + a_a^\dag a_b^\dag a_c).
\label{Hnl-abc}
\end{equation}
These processes are schematically shown in Fig.~\ref{Eqv-coupled-LCcircuits}.
In the case of SFC, the input signals have frequencies $\omega_a$ and $\omega_b$
around $\omega_1$, while the frequency of the output signal,
$\omega_c \approx \omega_2$.
In the case of SPDC, the second mode driven at frequency $\omega_c \approx \omega_2$,
gives rise to generating entangled photon pairs
within the first mode bandwidth. Thus, the double-mode circuit with $\chi^{(2)}$
nonlinearity can serve as a source of nonclassical microwave light, including
squeezed states  \cite{Movshovich1990}. Note that the three-photon SPDC was
recently observed in a microwave-driven superconducting cavity with the Kerr
nonlinearity \cite{Chang2020}.

\section{Conclusion}

We have demonstrated that the rf SQUIDs embedded in superconducting coplanar
waveguide resonators enabled the
observation of a number of remarkable parametric effects.
The degree of the non-centrosymmetric nonlinearity, enabling these effects,
is governed by the rf SQUID parameter $\beta_L$,
which value we extracted from resonant frequency modulation.
The asymmetric position of this nonlinear element, i.e., at one
third of the length of the open-ended resonator, allowed efficient
coupling of three waves within the two lowest modes, $n=1$ and $n=2$.
The range of observed 3WM processes include SHG, HTG, SFG, DFG, JPA
in both degenerate and non-degenerate modes of operation.

The obvious advantage of a resonator with nonlinearity of a $\chi^{(2)}$
type over its counterpart with the Kerr nonlinearity \cite{Tancredi2013}
is that the former circuit can provide a principally stronger coupling.
This property is particularly important for the open-ended resonators,
where the intermode coupling is proportional to $\alpha^2$ and $\alpha^3$,
respectively.

In conclusion, we believe that the superconducting cavities with
$\chi^{(2)}$ nonlinearity are suitable not only for experiments with
classical signals, but also for generating and converting nonclassical states
of microwave fields on the level of single photons. In perspective, these
circuits may enable nondemolition quantum measurements and quantum computing
operations without conventional qubits.
The proposed circuit will definitely extend the range of quantum
information experiments with microwave photons and thus will have impact
on quantum communication technologies.

\begin{acknowledgments}

The authors would like to thank J. Felgner, M. Petrich, T. Weimann, and
R. Gerdau for their help in fabrication of the samples and
M. Schr\"{o}der and V. Rogalya for improving the measuring setup.

This work has received funding from the EMPIR programme (project ParaWave
17FUN10) co-financed by the Participating States and from the European
Union's Horizon 2020 research and innovation programme, and from the
German Federal Ministry of Education and Research (funding programme
Quantum Technologies - from basic research to market, contract
number 13N15949).
CK acknowledges the funding of the Braunschweig International
Graduate School of Metrology B-IGSM and the DFG
Research Training Group 1952 Metrology for Complex Nanosystems.

\end{acknowledgments}

\appendix

\section{Resonant frequency of a cavity with embedded inductance}

A particular solution of a wave equation for the phase variable
$\phi(z,t)$ in the lossless open-ended resonator
(the coupling capacitances,
$C_\textrm{in}$ and $ C_\textrm{out}$, are assumed to be negligibly small),
having boundary conditions
\begin{equation}
\frac{\partial \phi(0,t)}{\partial z} = \frac{\partial \phi(d,t)}{\partial z}  = 0
\label{boundary}
\end{equation}
and embedded inductor $L_s$ (positioned not
necessarily in the center, as shown in Fig.~1) has the form of
a standing-wave,
\begin{equation}
\phi(z,t) = \psi(z) e^{- i\omega_nt}.
\label{phi_standing}
\end{equation}
Here $\omega_n$ is the frequency of the $n$-th mode and wave amplitude $\psi(z)$ is
a piecewise function whose values for the left and the right segments
of the transmission line are given by the harmonic (cosine) functions,
\begin{equation}
     \psi(z)=\left\{\begin{array}{ll} \psi_l(z)  = \psi_a \cos k_n z, & 0\leq z<x_l, \\
         \psi_r(z) = \psi_b \cos[ k_n (d - z)], & x_r< z\leq d. \end{array}\right.
\label{psi_standing1}
\end{equation}
Thus the outer boundary conditions Eq.~(\ref{boundary}) are fulfilled.

Wave number $k_n$ for mode $n = 1, 2, ...$ in the resonator with embedded
inductance $L_s$ is
\begin{equation}
k_n = k^{(0)}_n - \delta k_n,
\label{mode-kn}
\end{equation}
where the wave number of the bare resonator is  {\color{red}{\cite{Pozar1993}}}
\begin{equation}
k^{(0)}_n = n\pi /d
\label{mode-k0n}
\end{equation}
and $\delta k_n$ is the change of the wave number due to inductance $L_s$.
The continuous current conditions on the left and the right terminals of
inductance $L_s$ read
\begin{eqnarray}
\frac{1}{\ell} \psi_l'(x_l) = -\frac{1}{L_s} \left[\psi_l(x_l)
- \psi_r(x_r)\right] = \frac{\varphi}{L_s} ,  \label{cont-curr1} \\
\frac{1}{\ell} \psi_r'(x_r) = -\frac{1}{L_s} \left[\psi_l(x_l)
- \psi_r(x_r)\right]= \frac{\varphi}{L_s},
\label{contin-current2}
\end{eqnarray}
where $\varphi$ is the phase drop on inductance $L_s$. Using
Eq.~(\ref{psi_standing1}) the derivatives on the left-hand-sides of equations
(\ref{cont-curr1}) and (\ref{contin-current2}) are expressed as
\begin{equation}
\psi'_{l}(x_l) = - k_n \psi_a \sin k_n x_l
\label{derivative1}
\end{equation}
and
\begin{equation}
\psi'_{r}(x_r) =  k_n \psi_b \sin[ k_n (d - x_r)],
\label{derivative2}
\end{equation}
respectively.

Inserting relations (\ref{derivative1}) and (\ref{derivative2})
into Eqs. (\ref{cont-curr1})
and (\ref{contin-current2}) gives the set of two homogeneous
linear equations for wave amplitudes $\psi_a$ and $\psi_b$,
\begin{eqnarray}
a_{11} \psi_a + a_{12} \psi_b =0,  \label{eq1psi} \\
a_{21} \psi_a + a_{22} \psi_b =0. \label{eq2psi}
\end{eqnarray}
Here coefficients $a_{ij}$ ($i,j = 1,2$) are
\begin{eqnarray}
a_{11} &=& -\alpha k_n d \sin  k_n x_l + \cos  k_n x_l,  \label{a11} \\
a_{12} &=& - \cos [ k_n(d- x_r)], \label{a12}  \\
a_{21} &=&  \cos k_n x_l,  \label{a21} \\
a_{22} &=& \alpha k_n d \sin [ k_n(d- x_r)] -\cos [ k_n(d- x_r)], \,\,\,\,\,\,\,\,
\label{a22}
\end{eqnarray}
where the inductance participation ratio is
\begin{equation}
\alpha = L_s/\ell d = L_s/L_{\textrm{tot}}.
\label{alpha-def}
\end{equation}
The condition that the corresponding matrix has a zero determinant
and, hence, that a nonzero solution $(\psi_a, \psi_b )$ of
equations (\ref{eq1psi}) and (\ref{eq2psi}) exists,
\begin{equation}
\Delta = a_{11}a_{22}-a_{12}a_{21} = 0,
\label{det}
\end{equation}
yields the transcendent equation for wave number $k_n$,
\begin{equation}
\alpha k_n d \sin  k_n x_l \sin [ k_n(d- x_r)] =  \sin [ k_n (d-x_r +x_l)].
\label{eqFORkn}
\end{equation}

Taking into account the fact that both the participation ratio $\alpha$
and the relative size of inductor, $(x_r - x_l)/d =  \Delta x/d$ (see Fig. 1a),
are small $( \ll 1)$ and, thus, the wave-number change in
Eq.~(\ref{mode-kn})
is also small, $|\delta k_n| \ll n \pi/d$, Eq.~(\ref{eqFORkn}) is linearized.
The resulting simplified equation is
\begin{equation}
(-1)^{n+1}\alpha k_n^{(0)} d \sin^2 k_n^{(0)} x
= \sin ( n \pi -  \delta k_n d - k_n^{(0)} \Delta x)
\label{eq-kn-linearized}
\end{equation}
or
\begin{equation}
\alpha \sin^2 (n \pi x/d) =  \delta k_n /k_n^{(0)} +  \Delta x/d.
\label{eq-kn-linearized2}
\end{equation}
Thus the relative shift of the wave number and thus of the resonant frequency
for mode $n$ is
\begin{equation}
\frac{\delta k_n}{k_n^{(0)}} =-\frac{\omega_n -\omega^{(0)}_n}{\omega^{(0)}_n}
= \alpha^*_n \equiv \alpha \sin^2 \frac{n \pi x}{d} - \frac{\Delta x}{d}.
\label{delta-fnRes}
\end{equation}
Here the trigonometric factor $\sin^2 (n \pi x/d)$ gives the dependence on the inductor
position $x$, while term $ \Delta x/d$ describes the effective reduction of
the resonator length, $d \rightarrow d - \Delta x$.
The resulting eigenfrequencies are
\begin{equation}
\omega_n = \omega^{(0)}_n(1-\alpha_n^*),~~~~n = 1, 2, 3, ...
\label{ref-freq}
\end{equation}
and wave numbers are
\begin{equation}
k_n = n \pi d \, (1-\alpha_n^*).
\label{k-number}
\end{equation}
The corresponding eigensolutions of the homogeneous wave equation in the
range $0\leq z \leq d$ have the form
\begin{eqnarray}
\psi_n &=&  [1 - \alpha \pi n (-1)^{n} ]\,
\Theta(x_l-z)\cos k_n z \nonumber\\
&+&\Theta(z-x_r) \cos[ k_n (d - z)]    
\approx  \cos k_n z,
\label{eigensolution}
\end{eqnarray}
where $\Theta(z)$ is the Heaviside step function.

\section{Two-mode Hamiltonian of the cavity with $\chi^{(2)}$ nonlinearity}

The rf SQUID potential energy in the case of the optimal flux bias,
$\Phi_{\textrm{dc}} = \Phi_0/4$ (or $\varphi_{\textrm{dc}}
= \Phi_{\textrm{dc}}/\varphi_0 = \pi/2$),
and small ac phase difference on the Josephson junction, $|\varphi|\ll\pi/2$,
is given by
\begin{equation}
E_s = \varphi_0\int^\varphi I(\varphi) d\varphi = E_{s0}
\left(\frac{\varphi^2}{2} - \beta\frac{ \varphi^3}{3} \right),
\label{rfSQUIDenergy}
\end{equation}
where energy $E_{s0}=\varphi^2_0 /L_g$.
The first term ($\propto \varphi^2$) on the right hand side of this relation
contributes only to the eigenmode frequency Eq.~(\ref{ref-freq}).
Omitting, for a moment,  the nonlinearity of the rf SQUID, the Hamiltonian
of the resonator can be presented using the second quantization formalism
in the standard form,
\begin{equation}
H_0 =  \sum_{n = 1}^\infty \hbar \omega_n a_n^\dag a_n,
\label{Ham0}
\end{equation}
where $a^\dag_n$ and $a_n$ are creation and annihilation operators
of the $n$-th mode, respectively.

The last term on the right hand side of Eq.~(\ref{rfSQUIDenergy}),
that is
\begin{equation}
E_{\textrm{nl}}= -\frac{1}{3} \beta E_{s0} \varphi^3
= -\frac{1}{3} \beta \alpha^3 E_{s0} d^3 \, (\phi')^3,
\label{Enonlin}
\end{equation}
describes the energy associated with the rf SQUID nonlinearity.
Here the inductance participation ratio 
is $\alpha=L_g/L_\textrm{tot} \ll 1$ and phase $\varphi$ is expressed
via derivative $\phi'$
using Eqs.~(\ref{cont-curr1}) and (\ref{contin-current2}).
Using the normal-mode decomposition of phase $\phi$ \cite{Eichler2014},
\begin{equation}
\phi =\sum_{n=1}^\infty A_n \psi_n
= \sum_{n=1}^\infty A_n \cos k_n x,
\label{phi-n-decomp}
\end{equation}
the nonlinear part of the rf SQUID energy takes the form
\begin{equation}
E_{\textrm{nl}} = \sum_{l,m,n=1}^\infty B_{lmn} A_l A_m A_n,
\label{EnlDecomp}
\end{equation}
where
\begin{equation}
B_{lnm}  = \frac{1}{3} \beta \alpha^3 E_{s0}
 \prod_{j\in\{ l,m,n\}} k_j d \, \sin k_j x.
\label{Bjnm}
\end{equation}

In the quantum case, the coefficients in the normal-mode
decomposition Eq.~(\ref{phi-n-decomp})
are operators, $A_n \rightarrow \hat{A}_n$.
They can be expressed via annihilation ($a_n$) and
creation ($a^\dag_n$) boson operators \cite{Wallquist2006},
that is
\begin{equation}
\hat{A}_n  = \phi_{\textrm{zpf},n} (a^\dag_n + a_n),
\label{a+a}
\end{equation}
where the prefactor is associated with the magnitude
of zero-point fluctuations \cite{Eichler2014,Girvin2011},
\begin{equation}
\phi_{\textrm{zpf},n} =
\varphi_0^{-1}\sqrt{\hbar/2\omega_n C_{\textrm{eff},n}}.
\label{zpf-expression}
\end{equation}
Using the bare cavity values of frequencies $\omega_n$ (\ref{fRes0}) and
capacitances $C_{\textrm{eff},n}$ (\ref{Ceff}) we have
\begin{equation}
\phi_{\textrm{zpf},n} \approx \sqrt{\frac{2Z_0}{nR_Q}}\approx \frac{0.124}{\sqrt{n}},
\label{zpf-expression2}
\end{equation}
where impedance $Z_0 = \sqrt{L_\textrm{tot}/C_\textrm{tot}}=50~\Omega$ and resistance
quantum $R_Q = h/4e^2\approx 6.45~\textrm{k}\Omega$.
Inserting Eq.~(\ref{a+a}) in expression (\ref{EnlDecomp}) yields
nonlinear part of Hamiltonian $H_{\textrm{nl}}$ in terms of
operators $a_n$ and $ a^\dag_n$.

Applying a rotating wave approximation and thus
omitting all oscillating terms of type $a_l a_m a_n$,
$a^\dag_l a^\dag_m a^\dag_n$, etc., we keep only terms
$a_l a_m a^\dag_n$ and $a^\dag_l a^\dag_m a_n$ with the frequencies
obeying the three photon (3WM) relation,
\begin{equation}
\omega_l  + \omega_m -\omega_n =0,~~~\textrm{or},~~~l+m-n=0.
\label{3wmRelation}
\end{equation}
we obtain the total Hamiltonian, $H = H_0 + H_{\textrm{nl}}$, in the form
\begin{equation}
\frac{H}{\hbar} =  \sum_{n = 1}^\infty  \omega_n a_n^\dag a_n
+ \sum_{l+m-n=0}g_{lmn} ( a_l a_m a_n^\dag + a_l^\dag a_m^\dag a_n ),
\label{HamSum}
\end{equation}
where the mode coupling coefficient is
\begin{equation}
g_{lmn}= \frac{1}{\hbar}\left(\frac{2Z_0}{R_Q}\right)^{3/2}\sqrt{l m n} \, B_{lmn}.
\label{coupling-coeff}
\end{equation}
Assuming that relation (\ref{3wmRelation}) is fulfilled only for the two
lowest modes, i.e., $\omega_2 = 2 \omega_1$, Hamiltonian (\ref{HamSum}) takes
the form
\begin{equation}
H =  \sum_{n = 1}^2 \hbar \omega_n a_n^\dag a_n
+ \hbar g( a_1 a_1 a_2^\dag + a_1^\dag a_1^\dag a_2 ),
\label{Ham112}
\end{equation}
where
\begin{equation}
\hbar g= 3\hbar g_{112}= \beta \alpha^3 \left(\frac{3Z_0}
{R_Q}\right)^{3/2}E_{s0}
\label{coupling-coeff-g}
\end{equation}
or
\begin{equation}
\frac{g}{\omega_1} = \frac{3\beta \alpha^2}{2 \pi^2} \sqrt{\frac{3Z_0}
{R_Q}}.
\label{coupling-coeff-g-2}
\end{equation}
Here we used the relations $g_{112}=g_{121}=g_{211}\equiv g/3$ and
$\sin k_1 d = \sin k_2 d \approx \sqrt{3}/2$ and the commutative
property of the operators associated with different
modes, $[a_1,a_2^\dag] = [a_1^\dag,a_2]=0$.

\section{Sample measurements}

The samples were characterized using a vector network analyzer (VNA)
R$\&$S ZVA 40
by measuring transmission through the CPW resonator (parameters $S21$ and $b2$)
as a function of frequency and/or microwave power.
In the mixing experiments a microwave power supply was made using the
sources R$\&$S SMF100A,
while an output power was measured by a spectrum analyzer (R$\&$S FSV 30).
In the experiments in liquid helium, 
the chips were bonded on a
printed circuit board with the SMA connectors and mounted in a metallic box.
In the dilution fridge unit, the attenuators in the input rf line mounted
at different temperature stages gave an attenuation of $-50$~dB in total
(including attenuation in the cables).
In the output rf line, two circulators (with the bandwidth of $3.2-8.1$ GHz)
were installed at the mixing chamber level. The input and output rf lines
(coaxial cables) were directly bonded to the chip.
The flux bias lines were supplied with the on-chip low-pass
filters comprising the spiral-shape Nb coils with inductance about 5.5 nH.
In both setups the sample was protected by a cryoperm shield.
The readout of the output signal was done using
a semiconductor low-noise amplifier from Low Noise Factory,
model LNF-LNC0.3-14A. It operated in the range of 0.3-14 GHz with a gain
up to 40~dB and a noise temperature of about 5~K. This cryogenic amplifier was
installed at the 4~K stage of the dilution fridge unit.

\end{document}